\documentclass[twocolumn,showpacs,amsmath,prl,superscriptaddress]{revtex4-1}

\usepackage{color}
\usepackage{graphicx}
\usepackage{epsfig,psfrag}
\usepackage{amsmath}
\usepackage{amssymb}
\usepackage{amsbsy}
\usepackage{amsthm}
\usepackage{amsfonts}
\usepackage{wasysym}
\usepackage{bbm}
\usepackage{tabularx}
\usepackage{euscript}
\usepackage{color}
\usepackage{enumerate}
\usepackage{amsfonts}
\usepackage{exscale}
\usepackage{bbold}
\usepackage{float}
\usepackage{slashed}

\usepackage[colorlinks,citecolor=blue]{hyperref}

\newcommand{\be}{\begin{eqnarray}}
\newcommand{\ee}{\end{eqnarray}}

\newcommand{\Eq}[1]{equation~(\ref{#1})}
\newcommand{\Fig}[1]{Fig.~\ref{#1}}

\newcommand{\<}{\langle}
\renewcommand{\>}{\rangle}

\newcommand{\Tr}{{\rm Tr}}

\renewcommand{\Im}{{\rm Im}}
\renewcommand{\Re}{{\rm Re}}

\def\bea{\begin{eqnarray}}
\def\eea{\end{eqnarray}}
\def\bra#1{\left\langle#1\right|}
\def\ket#1{\left|#1\right\rangle}
\def\avg#1{\left\langle#1\right\rangle}

\def\Tr{\mathrm{Tr}}

\def\Eq#1{Eq.~(\ref{#1})}
\def\Fig#1{Fig.~\ref{#1}}

\begin{document}
\title{Deconfined quantum criticality and emergent SO(5) symmetry in fermionic systems}
\author{Zi-Xiang Li}
\affiliation{Department of Physics, University of California, Berkeley, CA 94720, USA}
\affiliation{Materials Sciences Division, Lawrence Berkeley National Laboratory, Berkeley, CA 94720, USA}
\author{Shao-Kai Jian}
\affiliation{Institute for Advanced Study, Tsinghua University, Beijing 100084, China}
\author{Hong Yao}
\affiliation{Institute for Advanced Study, Tsinghua University, Beijing 100084, China}
\begin{abstract}
{Deconfined quantum criticality with emergent SO(5) symmetry in correlated systems remains elusive. Here, by performing numerically-exact state-of-the-art quantum Monte Carlo (QMC) simulations, we show convincing evidences of deconfined quantum critical points (DQCP) between antiferromagnetic and valence-bond-solid phases in the extended Hubbard model of fermions on the honeycomb lattice with large system sizes. We further demonstrate evidences of the SO(5) symmetry at the DQCP. It is important to note that the critical exponents obtained by finite-size scaling at the DQCP here are consistent with the rigourous conformal bounds. Consequently, we established a promising arena of DQCP with emergent SO(5) symmetry in interacting systems of fermions. Its possible experimental relevances in correlated systems of Dirac fermions will be discussed briefly.}
\end{abstract}
\maketitle

Quantum criticality and emergent phenomena in correlated many-body systems are among central topics in modern condensed matter physics \cite{Sachdev-book,Sondhi-RMP}. Conventional quantum critical points (QCPs) may be described by the Landau-Ginzburg-Wilson (LGW) paradigm. It has been of great interest to fathom QCPs beyond the LGW paradigm \cite{Sachdev-book,Sondhi-RMP,Wen-book,Fradkin-book}. One prototype non-Landau QCP is deconfined quantum critical point (DQCP) \cite{Senthil-04a,Senthil-04b}, which is a continuous transition between certain ordered states and where discontinuities would appear in conventional LGW paradigm. Fractionalized excitations such as deconfined spin-1/2 spinons appear at DQCPs although they are confined in ordered phases. Tremendous progress has been made in identifying and understanding DQCPs in quantum magnets \cite{Levin-Senthil-04, Motrunich-04,Hu-05,Senthil-06,Lee-10, Jian-18,You-18,Motrunich-19a, Sandvik-07,Kaul-08, Wiese-08,Prokofev-08,Kaul-12,Xiang-12, Alet-13,Kaul-13, FWang-15, Guo-16,Meng-18a,Meng-18b,Pollmann-18, Max-18,Gazit-18,Xiang-19,Motrunich-19b,Sreejith-15,Harada-13,Alet-15,Sandvik-09,Kaul-132,Prokofev-13,Bartosch-13,Wen-16}.

Enlarged symmetry often emerges in low-energy at QCPs. For a DQCP \cite{Senthil-04a,Senthil-04b} separating two ordered phases of antiferromagnetism (AFM) and valence-bond-solid (VBS) in SU(2)-invariant quantum magnets \cite{Haldane-88,Read-Sachdev-89}, it was argued from duality relations that SO(5) symmetry may emerge in low-energy 
\cite{Senthil-15,Metlitski-16,Wang-17,Wang-16,Meng-17,Cenke-18}. The emergent SO(5) symmetry can unify VBS and AFM such that one order can be rotated into the other \cite{Zhang-97}. Such emergent SO(5) symmetry enriches physics of DQCPs; for instance, assuming SO(5) symmetry at DQCPs, rigorous bounds of certain critical exponents can been obtained from conformal bootstrap calculations \cite{Bootstrap-16,Bootstrap-RMP19}. Evidences of emergent SO(5) symmetry were reported in simulations of 3D classical loop model that exhibits a putative DQCP \cite{Nahum-15a,Nahum-15b}. Nevertheless, critical exponents obtained in previous studies are not entirely consistent with the rigorous conformal bounds, raising concerns whether DQCP with emergent SO(5) symmetry really occurs there \cite{Bootstrap-RMP19}. Consequently, it is highly desired to explore DQCPs with emergent SO(5) symmetry and with critical exponents consistent with the conformal bounds.

\begin{figure}[t]
\includegraphics[height=4.cm]{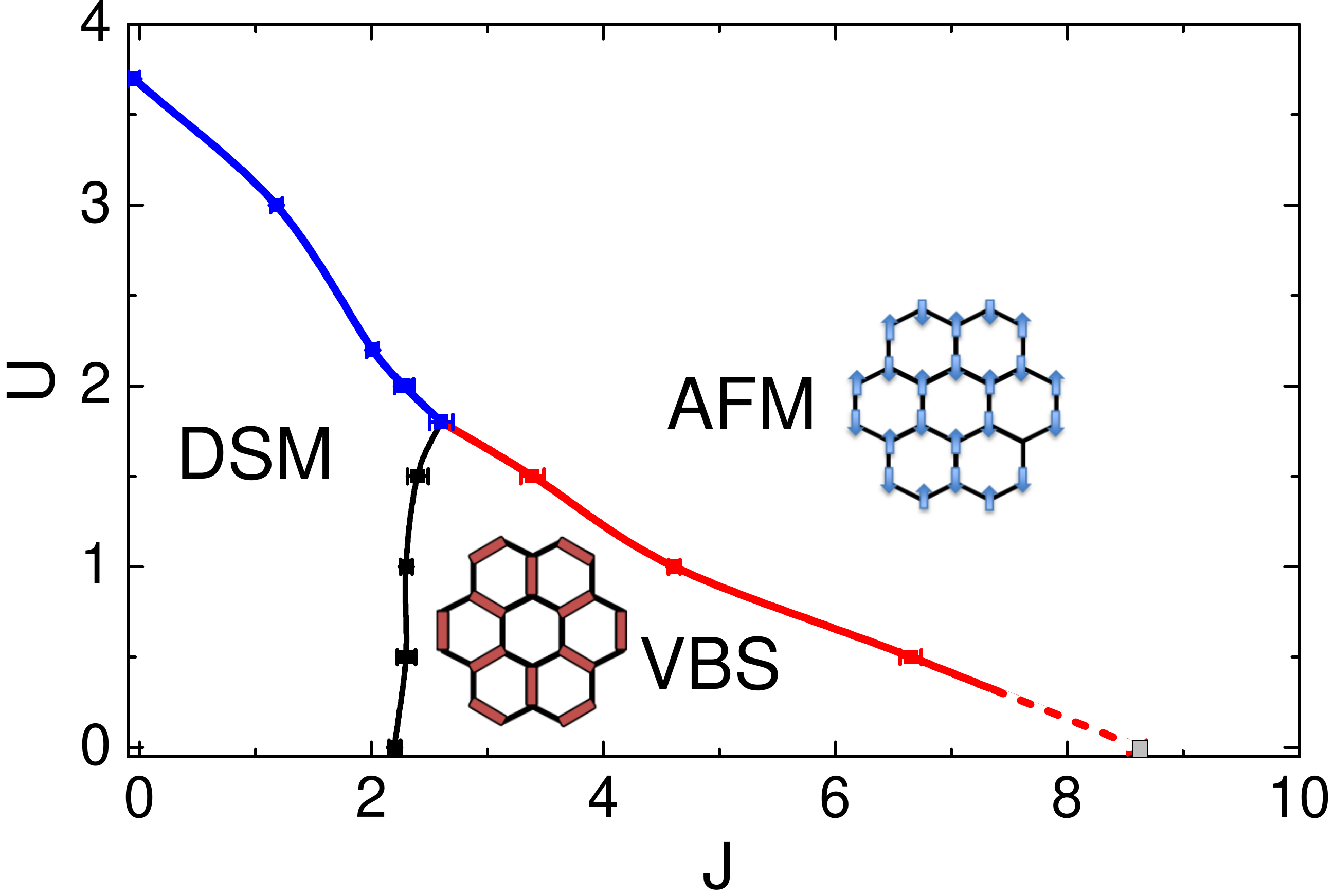}
\caption{The quantum phase diagram of the extended Hubbard model on the honeycomb lattice is obtained from large-scale QMC simulations. Here $U$ and $J$ label the strength of Hubbard and effective SSH interactions, respectively. It is remarkable that the continuous quantum phase transition (solid line) between the VBS and AFM phases features deconfined quantum critical point with emergent SO(5) symmetry.}
\label{fig1}
\end{figure}

Here we investigate a microscopic model of fermions on the honeycomb lattice which describes a Dirac semimetal (DSM) at weak interactions but exhibits a direct quantum phase transition (QPT) between VBS and AFM phases for strong interactions, as shown in Fig.~\ref{fig1}. Remarkably, this model is free from notorious fermions-sign problem \cite{Troyer-05, Wu-05, LJY-15, LJY-16, Wang-15, Xiang-16} (see Ref. \cite{LY-18} for a recent review) and is therefore amendable to large-scale numerically-exact quantum Monte Carlo (QMC) simulations \cite{BSS-81,Sorella-89,White-89,Assaad-05, Cerperley-86,Sandvik-98,Prokofev-98,Gull-11}. By performing state-of-the-art simulations of the model with large system size, we show convincing evidences that the direct VBS-AFM transition is continuous, namely featuring a DQCP. More importantly, the critical exponents extracted from finite-size scaling (FSS) analysis of QMC results are consistent with the rigorous conformal bounds. We further show evidences of emergent SO(5) symmetry at the DQCP by demonstrating the invariance of continuous rotations between AFM and VBS orders. To the best of our knowledge, it is the \textit{first} example of DQCP with both emergent SO(5) symmetry and critical exponents obeying the conformal bounds. We believe that it will provide a promising platform to investigate exotic physics of DQCPs and shed new light to emergence of higher symmetry in correlated systems such as interacting graphene-like materials with Kekul\'e VBS order \cite{Gutierrez-16}.

{{\bf Model:}} To investigate the QPT between VBS and AFM phases, we consider the following microscopic model of spin-1/2 fermions on the honeycomb lattice:
\bea
&&H=-t\sum_{\avg{ij}\sigma} (c^\dagger_{i\sigma}c_{j\sigma} + H.c.) +U\sum_i(n_{i\uparrow}-\frac{1}{2})(n_{i\downarrow}-\frac{1}{2}) \nonumber\\
&&~~~~~~~~~~~~~~~~~~~~~~~~~~
-\frac{J}{4} \sum_{\avg{ij}} \Big(\sum_\sigma c^\dagger_{i\sigma}c_{j\sigma} + H.c.\Big)^2,
\label{model}
\eea
where $c^\dag_{i\sigma}$ creates a fermion with spin polarization $\sigma=\uparrow$,$\downarrow$ on site $i$, $n_{i\sigma}=c^\dag_{i\sigma}c_{i\sigma}$, and $\avg{ij}$ labels nearest-neighbor sites. The non-interacting part in \Eq{model} describes a Dirac semimetal on the honeycomb lattice. While $U$ represents the onsite Coulomb (Hubbard) interaction, the $J$ term can be considered as an effective Su-Schrieffer-Heeger (SSH) interaction as it can be induced by SSH electron-phonon couplings in the fast phonon limit \cite{SSH,Fradkin-83,Assaad-SSH}. At half-filling, namely the Fermi level lies at the Dirac point, the Hubbard $U$ favors AFM ordering while the SSH interaction $J$ favors VBS order \cite{LJJY-17}. We expect that a QPT between VBS and AFM phases should occur by varying the interactions $U$ and $J$ in the model above.

To stabilize VBS order in a larger parameter regime, we consider an additional interaction in \Eq{model}: $H_p \!=\! -\frac{P}4 \sum_{\alpha}( B_{\alpha}^2 + \tilde{B}_{\alpha}^2)$, where $B_{\alpha} \!=\!\sum_\sigma(c^\dagger_{\alpha_1\sigma}c_{\alpha_2\sigma} \!+\!c^{\dagger}_{\alpha_3\sigma}c_{\alpha_4\sigma} \!+\!c^\dagger_{\alpha_5\sigma}c_{\alpha_6\sigma} \!+\! H.c.)$ and $\tilde B_{\alpha} \!=\! \sum_\sigma(c^\dagger_{\alpha_1\sigma}c_{\alpha_6\sigma} \!+\!c^{\dagger}_{\alpha_5\sigma}c_{\alpha_4\sigma} \!+\!c^\dagger_{\alpha_3\sigma}c_{\alpha_2\sigma} \!+\! H.c.)$ are operators on the hexagon plaquette $\alpha$ consisting of six sites $\alpha_1$,$\cdots$,$\alpha_6$ arranged clockwise. Such plaquette interaction favors Kekul\'e VBS ordering and is expected to enlarge the region of VBS in the phase diagram. It is important to note that the model in \Eq{model} with finite $P$ is still free from the fermion-sign problem in QMC simulations; consequently we can perform unbiased and numerically-exact simulations of the model with large system sizes. Hereafter, we set $t\!=\!1$ and $P\!=\!1/4$ to explore the quantum phase diagram by varying interactions $U$ and $J$. We employ projector QMC \cite{Sorella-89,White-89,Assaad-05} simulations to investigate the ground-state properties of the model as a function of $U$ and $J$. Details of projector QMC can be seen in Supplemental Materials (SM).

\begin{figure}[t]
\includegraphics[height=2.65cm]{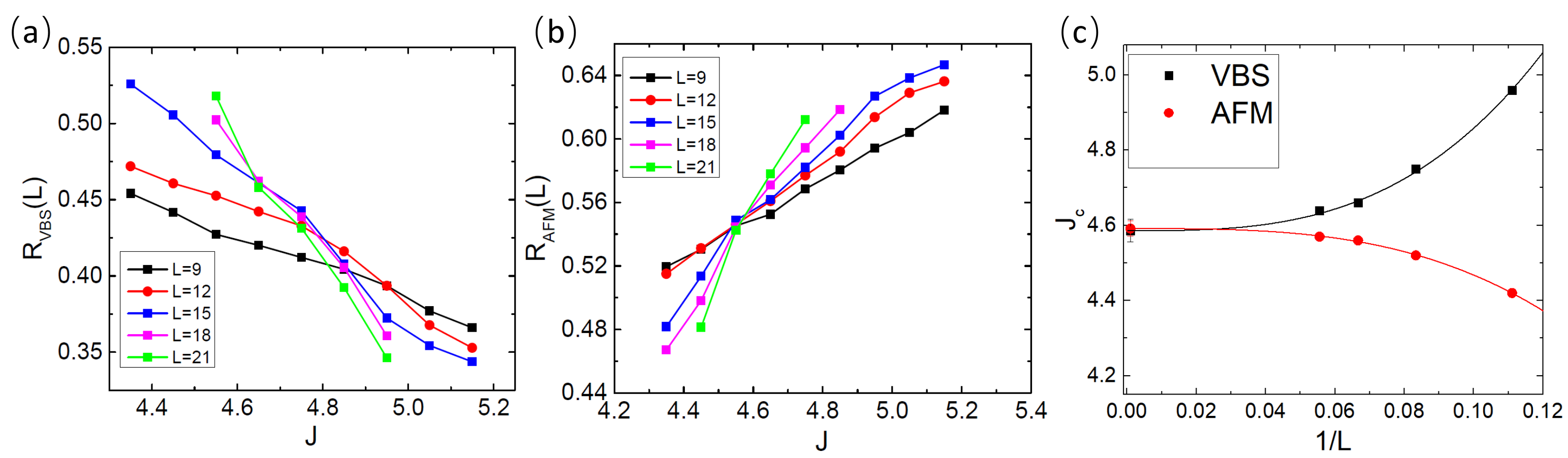}
\caption{(a) The QMC results of RG-invariant ratio for the VBS order at $U=1.0$. The largest system size in our simulation is $L=21$. (b) The results of RG-invariant ratio for the AFM order at $U=1.0$. (c) The critical values $J_c$ for the VBS or AFM phase transitions are extrapolated to the thermodynamic limit. The  transition points of VBS and AFM ordering are $J_{\textrm{VBS}}^c \!=\! 4.58\pm0.03$ and $J_{\textrm{AFM}}^c \!=\! 4.59\pm 0.02$. That $J_{\textrm{VBS}}^c$ and $J_{\textrm{AFM}}^c$ are identical within error bar indicates a direct transition between the VBS and AFM phases.}
\label{fig2}
\end{figure}

{{\bf Quantum phase diagram:}} To determine the boundary of VBS or AFM phases, we compute the RG-invariant ratio of VBS or AFM order parameters by QMC: $R_{\textrm{VBS/AFM}}\! =\! 1 - \frac{S_{\textrm{VBS/AFM}}(\vec{Q}+\delta \vec{q},L)}{S_{\textrm{VBS/AFM}}(\vec{Q},L)}$, where $S_{\textrm{VBS/AFM}}(\vec{q},L)$ is the structure factor of the VBS (AFM) on the honeycomb lattice with $2L^2$ sites. Here $\vec{Q}$ labels the momentum at which the structure factor is maximum; $\vec Q=\vec K$ for the VBS order while $\vec Q=\Gamma$ point for AFM order. $\delta \vec{q}$ is minimal momentum shift from $\vec Q$. Due to scaling invariance at QCPs, good crossing of the curves of $R$ as a function of tuning parameter $J$ for different system size $L$ would indicate continuous emergence of the order parameter. In our QMC simulations, we vary the SSH interaction $J$ to find phase boundaries while fixing the Hubbard interaction $U$.

The quantum phase diagram obtained from large-scale QMC simulations is shown in \Fig{fig1}. For relatively weak Hubbard interaction $U\!<\!U_0$ ($U_0\!\approx\!1.8$), we find that the system exhibits three qualitatively different phases as a function of $J$. When $0\!<\!J\!<\!\tilde J_c$, the ground state is a DSM where neither VBS nor AFM ordering appears. The system is in the VBS phase when $J$ is increased across $\tilde J_c$. The DSM-VBS transition at $J\!=\!\tilde J_c$ is continuous despite the $Z_3$ nature of the VBS order-parameter and was dubbed as fermion-induced quantum critical points \cite{LJJY-17,Jian-17a,Herbut-17}. When $J$ is further increased across $J_c$, VBS ordering vanishes while AFM emerges. The accurate phase boundaries of VBS and AFM can be determined from the crossing point of RG-invariant ratio $R_\textrm{VBS/AFM}$. The details of extracting transition points and of extrapolating to thermodynamic limit are summarized in the SM. For $U=1.0$, the RG-invariant ratios of the VBS and AFM order parameters are shown in \Fig{fig2}(a) and \Fig{fig2}(b), respectively. The extrapolation of critical values $J_c(L)$ for VBS or AFM orderings to the thermodynamic limit $L\!\to\!\infty$ is shown in \Fig{fig2}(c). It convincingly shows that, after extrapolating to the thermodynamics limit, the critical value of $J$ at which the VBS ordering vanishes is identical with the one where the AFM order emerges within error bar. This indicates that for $U\!=\!1.0$ the system undergoes a direct QPT from the VBS to AFM phases when $J$ is varied. The direct VBS-AFM transitions as a function of $J$ are also observed for $U=0.5$ and $U=1.5$, as shown in the SM.

As the Hubbard interaction suppresses VBS but favors AFM ordering, the intermediate $J$ region with VBS order shrinks upon increasing $U$, as shown in \Fig{fig1}. For intermediate $U$, namely $\tilde U_0\!>\!U\!>\!U_0$ ($\tilde U_0\!\approx\! 3.7$), the VBS phase does not appear and system undergoes a direct transition from DSM to AFM as $J$ is varied. The DSM-AFM transition is usually continuous and should be in the Gross-Neveu-Heisenberg universality class \cite{Assaad-13,Sorella-16,Assaad-15,Herbut-17b}. Indeed, the critical exponents at the DSM-AFM transition obtained in our QMC simulations, as shown in the SM, are consistent with previous studies \cite{Assaad-13,Sorella-16,Assaad-15}. There is a multicritical point where three phases (namely DSM, VBS and AFM) meet. For strong $U$ ($U\!>\!\tilde U_0$), the system is in AFM phase for all $J$.

In the following, we shall focus on the direct transition between the VBS and AFM phases observed in our QMC simulations, which is putatively a first-order transition in the LGW paradigm. Especially, we would like to investigate if the direct transition could be a DQCP beyond the LGW paradigm and if an enhanced symmetry such as SO(5) can emerge at the transition.

\begin{figure}[t]
\includegraphics[height=2.6cm]{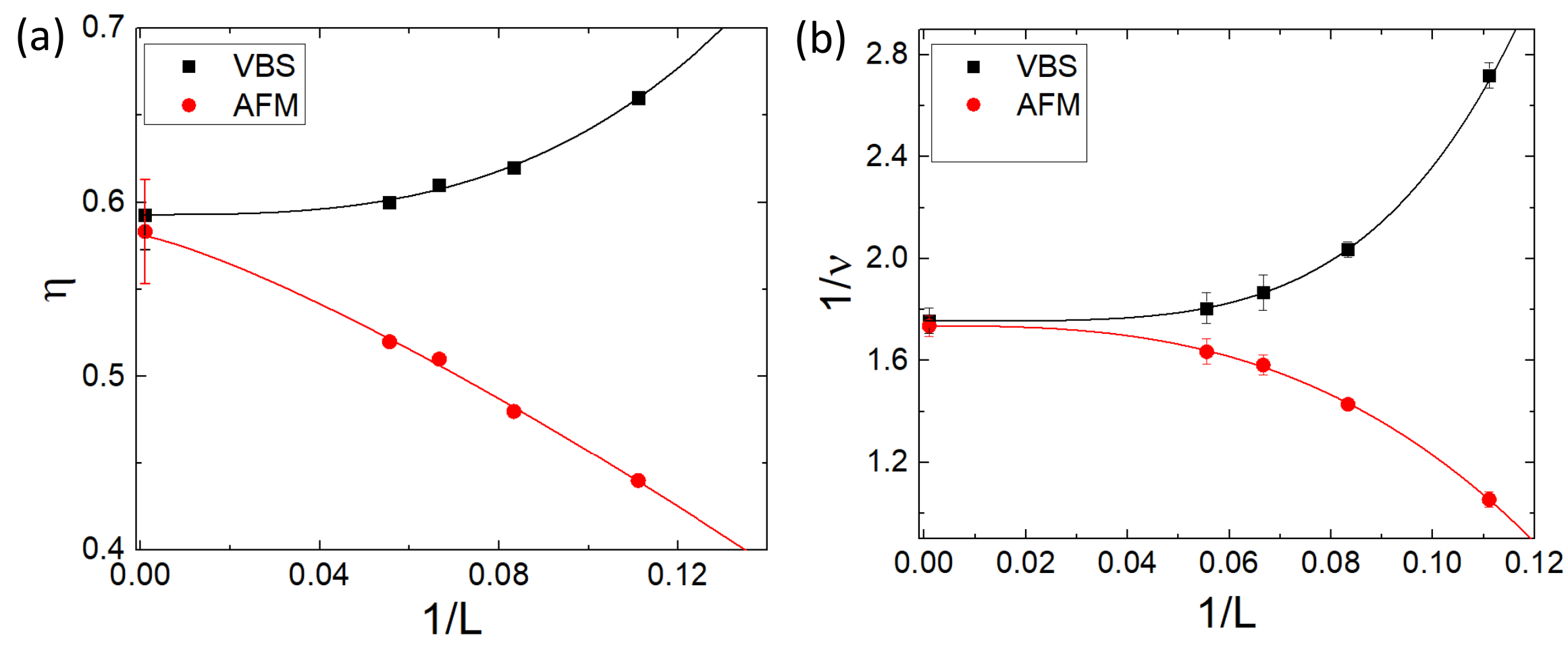}
\caption{The critical exponents for VBS and AFM orders at $U\!=\!1.0$ obtained by finite-size scaling: (a) Anomalous dimensions: $\eta_{\textrm{VBS}}\!=\!0.59 \pm 0.02$, $\eta_{\textrm{AFM}}\!=\!0.58\pm0.03$; (b) Correlation length exponent: $1/\nu_{\textrm{VBS}} \!=\! 1.76\pm 0.05$, $1/\nu_{\textrm{AFM}}\!=\!1.74\pm 0.03$. Both exponents are consistent with conformal bounds.}
\label{fig3}
\end{figure}

{{\bf Deconfined quantum critical point:}} To investigate whether DQCP occurs at the direct VBS-AFM transition, we first compute the first-order derivative of the ground-state energy with respect to $J$ around the transition. If a sharp kink in the derivative appears at the transition, it would indicate a first-order transition. For both $U\!=\!1.0$ and $U\!=\!0.5$, our QMC simulations show that no sharp kink appears even for the largest studied system size of $L\!=\!21$. It implies that the VBS-AFM transition could be continuous.

To better answer the question whether the direct VBS-AFM transition is first-order or continuous, we perform FSS analysis to extract the putative critical exponents of the transition. If the VBS-AFM transition is described by a DQCP, critical exponents of the two order parameters should be equal. We first obtain the anomalous dimension $\eta(L)$ and the correlation-length exponent $\nu(L)$ of order-parameter field for a given system size $L$, and then extrapolate their values to the thermodynamics limit by FSS. The details of computing critical exponents $\eta(L)$ and $\nu(L)$ are given in the SM. The obtained results of critical exponents $\eta$ and $\nu$ for the VBS and AFM orders at $U\!=\!1.0$ are shown in \Fig{fig3}(a) and (b), respectively. After FSS analysis, we obtain the critical exponents as follows: $\eta_{\textrm{VBS}} \!=\! 0.59\pm0.02$ and $\nu_{\textrm{VBS}} \!=\! 1.76\pm0.05$ for VBS while $\eta_{\textrm{AFM}}\!=\!0.58\pm0.03$ and $\nu_{\textrm{AFM}}\!=\!1.74\pm0.03$ for AFM. Intriguingly, the critical exponents for the VBS and AFM orders are equal within error bars, which provides a strong evidence that the VBS-AFM transition is continuous and is a DQCP. We further investigate the VBS-AFM transition for $U\!=\!0.5$ and the details are in the SM. The critical exponents of VBS and AFM orders for $U\!=\!0.5$ are also equal to each other within error bars and are consistent with the results for $U\!=\!1.0$, showing that the VBS-AFM transitions in the model are described by a DQCP of the same universality class.

\begin{figure}[t]
\includegraphics[height=2.5cm]{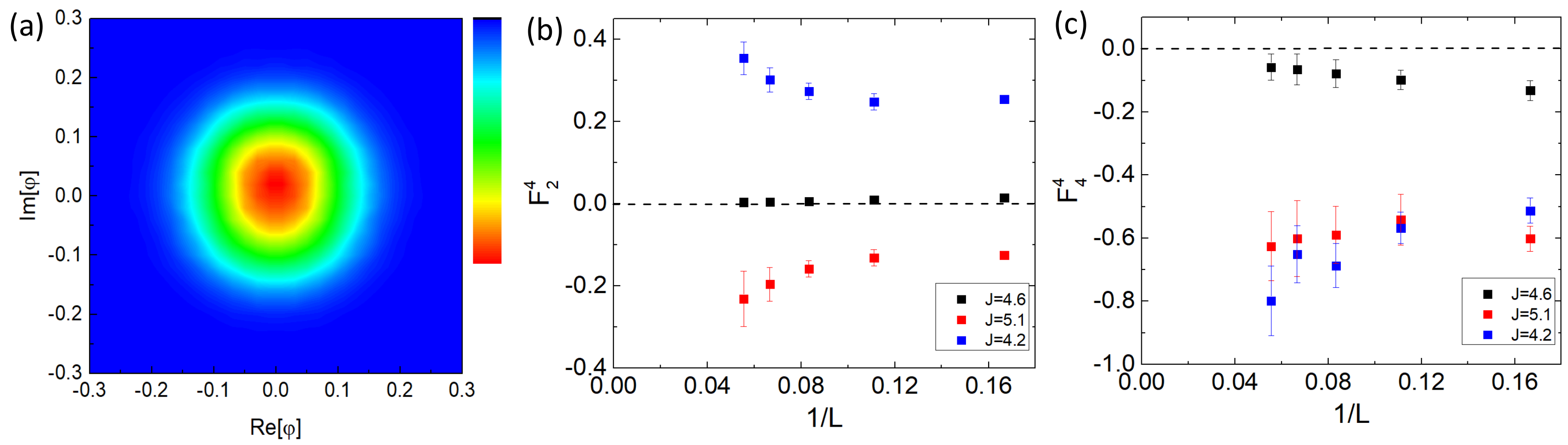}
\caption{(a) The concurrence probability $P(\Re[\phi],\Im[\phi])$ of VBS order parameter at VBS-AFM transition point ($J\!=\!4.6$ for $U\!=\!1.0$ and $L\!=\!15$). (b) The results of order parameter moment $F_2^4$ at and away from VBS-AFM phase transition point. (c) The results of order parameter moment $F_4^4$ at and away from VBS-AFM phase transition point.}
\label{fig4}
\end{figure}

Our simulations show that single-particle fermionic excitations are always gapped at and in the vicinity of the VBS-AFM transition (see the SM for details). Consequently, the VBS-AFM transition here is bosonic in nature although the system is composed of interacting fermions. The DQCP at the transition should be presumably described by the non-compact CP$^{1}$ theory \cite{Senthil-04a,Senthil-04b} where the operators creating or annihilating three monopoles of emergent gauge fields should be irrelevant at the transition \cite{Kaul-13} and U(1) rotational symmetry should emerge. Indeed, as shown in \Fig{fig4}(a), the concurrence probability of VBS order parameter is rotational invariant, implying the enlarged U(1) symmetry from $Z_3$ at the VBS-AFM transition. Consequently, we have shown convincing evidences that the direct VBS-AFM transition is a DQCP where U(1) symmetry emerges.

{\bf Emergent SO(5) symmetry:} Near the VBS-AFM transition, integrating out fermions would give rise to an effective field theory of five-component normalized order-parameter fields $\Phi\propto(\phi^1_\textrm{AFM}, \phi^2_\textrm{AFM}, \phi^3_\textrm{AFM},\phi^1_\textrm{VBS}, \phi^2_\textrm{VBS})$ with $|\Phi|=1$. The effective field theory is the 2+1D non-linear Sigma model (NLSM) with the following topological Wess-Zumino-Witten (WZW) term \cite{Senthil-06,Hu-05,Lee-10}:
\bea
S_\textrm{WZW}[\Phi]\!=\!\frac{i2\pi\epsilon_{abcde}}{\textrm{Area}(S^4)} \!\int\!dt\!\int\!d^3x \Phi^a \partial_t\Phi^b \partial_\tau \Phi^c \partial_x \Phi^d \partial_y \Phi^e,~~~~
\eea
where $x^\mu\!=\!(\tau,x,y)$, $t\!\in\![0,1]$ is the auxiliary parameter introduced to extend the field $\Phi(x^\mu)$ to $\Phi(t,x^\mu)$ with $\Phi(0,x^\mu)=\Phi(x^\mu)$ and $\Phi(1,\mu)=(0,0,0,0,1)$, and Area$(S^4)=\frac{8}{3}\pi^2$ labels the area of unit sphere $S^4$. The NLSM model with the topological WZW term suggests that the DQCP may feature emergent SO(5) symmetry, namely the invariance under rotations between VBS and AFM orders. Moreover, it was recently argued that the duality web associated with the critical NCCP$^1$ theory requires emergent SO(5) symmetry at the DQCP. The SO(5) symmetry is realized anomalously at the DQCP in the sense that it cannot occur at VBS-AFM transition in models which explicitly respect SO(5) symmetry \cite{Wang-17}.

Certain evidences of emergent SO(5) symmetry were reported in recent studies of DQCPs at VBS-Neel transitions \cite{Nahum-15a,Nahum-15b}. However, the order-parameter anomalous dimension $\eta\approx 0.2$$\sim$$0.4$ is obtained in all previous studies \cite{Sandvik-07,Kaul-08,Alet-13, Sandvik-09,Kaul-132,Nahum-15a, Nahum-15b,Guo-16,Guo-18,Zaletel-18} is not consistent with the quasi-rigorous bounds $\eta\!>\!0.52$ and $1/\nu \!<\!1.957$ obtained from conformal bootstrap calculation assuming the enlarged SO(5) symmetry at DQCP \cite{Bootstrap-RMP19}, indicating that the previously reported emergence of SO(5) symmetry should be a consequence of intermediate length scale and may be broken in the thermodynamic limit. Note that the critical exponents $\eta\!\approx\!0.59$ and $1/\nu\!\approx\!1.76$ obtained in the present large-scale QMC study are totally \textit{consistent} with the conformal bounds. This consistency suggests that SO(5) symmetry can emerge in the thermodynamic limit at the DQCP between VBS and AFM phases.

To further verify if SO(5) symmetry emerges at DQCP quantitatively, we compute the expectation value of operators which are not invariant under SO(5) rotations. If expectation values of all non-SO(5)-invariant operators vanish, it would support the emergence of SO(5) symmetry. One natural choice of non-SO(5)-invariant quantity is the $l\!\neq\!0$ moment $F_l^a = \avg{r^a \cos(l \theta)}$, where $r\cos\theta $ and $r\sin\theta$ represent normalized AFM and VBS order parameters, respectively \cite{Nahum-15a}. In the Ginzburg-Landau field theory describing the VBS-AFM transition, the lowest-order possible SO(5) anisotropy is from the quartic terms of order parameters, whose expectation values are the moments $F_2^4$ and $F_4^4$. Therefore, we compute the expectation values of $F_2^4$ and $F_4^4$ which are shown in \Fig{fig4}(b) and (c), respectively. It shows that the moments $F_2^4$ and $F_4^4$ vanish within error bar at the VBS-AFM transition in systems with large size. Higher-order moments, corresponding to the expectation values of higher-order terms in the Ginzburg-Landau field theory should be more irrelevant in the sense of RG. Combining the results of vanishing moments and critical exponents consistent with the rigorous conformal bounds, we conclude that the direct VBS-AFM transition in the extended Hubbard model features a DQCP with emergent SO(5) symmetry.

When Hubbard interaction $U$ is tuned to a special value $U\!=\!U_0$, the DSM-VBS transition line and VBS-AFM transition line cross, namely realizing a multicritical point where three phases of DSM, VBS and AFM meet \cite{Herbut-18,Roy-18,Assaad-17b}. It is argued from perturbative RG calculations that in the presence of low-energy Dirac fermions this multicritical point features an emergent SO(5) symmetry \cite{Herbut-18,Roy-18}. In other words, the SO(5) symmetry emerges at both the VBS-AFM transition line which ends at the DSM-VBS-AFM multicritical point.

{{\bf Discussions and concluding remarks:}} As the model in \Eq{model} with $U\!>\!0$ can be mapped into one with $U\!<\!0$ through the spin-particle-hole transformation, the quantum phase diagram with $U\!<\!0$ can be directly obtained from the one with $U\!>\!0$ through the transformation. While the VBS order is invariant under the transformation, the spin AFM order is transformed into the pseudospin-AFM order consisting of superconductivity (SC) and charge-density-wave (CDW) \cite{Zhang-90,Yang-90}. In other words, the ground state in the corresponding $U\!<\!0$ region is degenerate between SC and CDW, where SC and CDW can be rotated to each other through the pseudospin SO(3) transformation.

For the special case of $U\!=\!0$, the model preserves the spin-particle-hole symmetry such that the ground state with AFM order is degenerate to the one with pseudospin-AFM (namely SC/CDW) order. For this case, our QMC simulations still show a direct transition between VBS and AFM (or pseudospin-AFM) phases. However, due to the extra spin-particle-hole ($Z_2$) symmetry at $U\!=\!0$, the low-energy field theory describing the transition between VBS and AFM (pseudospin-AFM) phases should be qualitatively different from the usual NLSM with \textit{one} WZW term. By integrating out fermions, the resulting field theory in terms of order-parameter fields is the NLSM with {\it two} different WZW terms; one WZW term $S_\textrm{WZW}[\Phi]$ comes from the normalized five-component order-parameter vector $\Phi\!\propto\!(\phi^1_{\textrm{AFM}},\phi^2_{\textrm{AFM}},\phi^3_{\textrm{AFM}},\phi^1_{\textrm{VBS}}, \phi^2_{\textrm{VBS}})$ and the other $S_\textrm{WZW}[\bar \Phi]$ from a different normalized five-component order-parameter vector ${\bar\Phi}\!\propto \!(\phi^1_{\textrm{SC}}, \phi^2_{\textrm{SC}}, \phi_{\textrm{CDW}}, \phi^1_{\textrm{VBS}}, \phi^2_{\textrm{VBS}})$ (see the SM for details). Note that $\Phi$ and $\bar \Phi$ are not independent as both of them have the same field $\phi_\textrm{VBS}$; namely the two WZW terms are entangled. The precise physics of the NLSM model with two entangled WZW terms is unknown and is presumably different from the one with only one WZW term describing the conventional VBS-AFM DQCP.

Large-scale QMC simulations reveal evidences that the direct transition between the VBS and AFM (SC/CDW) phases at $U\!=\!0$ is first-order. Firstly, the exponents $\nu$ of VBS and AFM orders are extrapolated to small values, even smaller than $1/3$, as shown in the SM. The result of $\nu\!<\!1/d$, where $d=3$ represents the spacetime dimensions in the present study, is a strong indication of first-order transition. Secondly, the finite-size scaling results of the anomalous dimensions at the transition with $U\!=\!0$ are apparently different between VBS and AFM orders, which also indicates that the transition is not a DQCP. Putting together, when $U\!=\!0$ the direct transition between VBS and AFM (SC/CDW) in the present model is first-order in nature. As the first-order transition is expected to extend over a finite range of $U$ (the dashed line in \Fig{fig1}), there should be a tricritical point connecting the first-order and second-order phase transition boundaries between VBS and AFM. At the first-order transition between AFM and VBS, it is possible that enlarged SO(5) symmetry can emerge there due to the presence of WZW term \cite{Kaul-18,Sandvik-18}. Further understanding of the NLSM theory with \textit{two} entangled WZW terms at $U\!=\!0$ will help understand the first-order transition, which is left for future works.

In conclusion, we have proposed an interacting fermionic model on the honeycomb lattice which hosts intertwined orders at relatively strong interactions \cite{Kivelson-15} and exhibits direct quantum phase transition between VBS and AFM phases. From state-of-the-art unbiased QMC calculations, we have shown that the VBS-AFM transition is consistent with the paradigm of DQCP. More intriguingly, the large-scale simulations provide convincing evidences that an enlarged SO(5) symmetry emerges at the DQCP, which shed new light on understanding of exotic quantum critical points in strongly correlated electronic systems. The critical exponents obtained at the DQCP are consistent with the conformal bounds, providing a convincing arena of DQCP with emergent SO(5) symmetry. Finally, we would like to mention that the effective interactions which can drive the VBS-AFM transition can be induced by SSH electron-phonon coupling \cite{SSH}. Therefore, our work could not only shed new light to realizing emergent symmetry in correlated fermion systems but also pave an important step to the realizing DQCP in quantum matters such as graphene-like materials where the Kekul\'e VBS ordering has been observed in scanning tunneling measurements \cite{Gutierrez-16}.

{\it Acknowledgement}: We thank Chao-Ming Jian, Dung-Hai Lee, T. Senthil, Chong Wang, and Cenke Xu for helpful discussions. This work was supported in part by the MOSTC under grant Nos. 2016YFA0301001 and 2018YFA0305604 (HY), the NSFC under grant 11825404 (SKJ and HY), and the Gordon and Betty Moore Foundation EPIC Initiative under grant GBMF4545 (ZXL).

\begin{widetext}
\section{Supplementary Materials}

\renewcommand{\theequation}{S\arabic{equation}}
\setcounter{equation}{0}
\renewcommand{\thefigure}{S\arabic{figure}}
\setcounter{figure}{0}
\renewcommand{\thetable}{S\arabic{table}}
\setcounter{table}{0}

\subsection{I. The details of projector Quantum Monte Carlo simulation}

We perform projector QMC simulations to investigate the ground-state properties of the model introduced in the main text. We evaluate the ground-state expectation values of observables according to $\langle {\hat{O}}\rangle =\bra{\psi_0} \hat{O} \ket{\psi_0}/\<\psi_0 \mid \psi_0\> = \lim_{\theta\rightarrow \infty} \<\psi_T\mid e^{-\theta H } \hat{O} e^{-\theta H} \mid\psi_T\>/\<\psi_T\mid e^{-2\theta H}\mid \psi_T\>$, where $|\psi_T\>$ is a trial wave function. In our work, we choose $|\psi_T\>$ as the ground-state wave function of the non-interacting part of the model under study. Here $\theta$ is projection parameter which should be large enough to access the ground-state information. In our simulations, we choose $\theta=36/t$ for relatively small systems ($L\leq 12$) and $\theta=60/t$ for relatively large systems ($L>12$). We have checked the convergence of results against using larger values of $\theta$ and results obtained converge when larger $\theta$ is used. In projector QMC, we perform Trotter decomposition to discretize the imaginary time and the time step is $\Delta\tau = 0.05/t$ in our simulations. We also check that $\Delta\tau$ is small enough to guarantee the convergence of the results.

\begin{figure}[b]
\includegraphics[height=3.5cm]{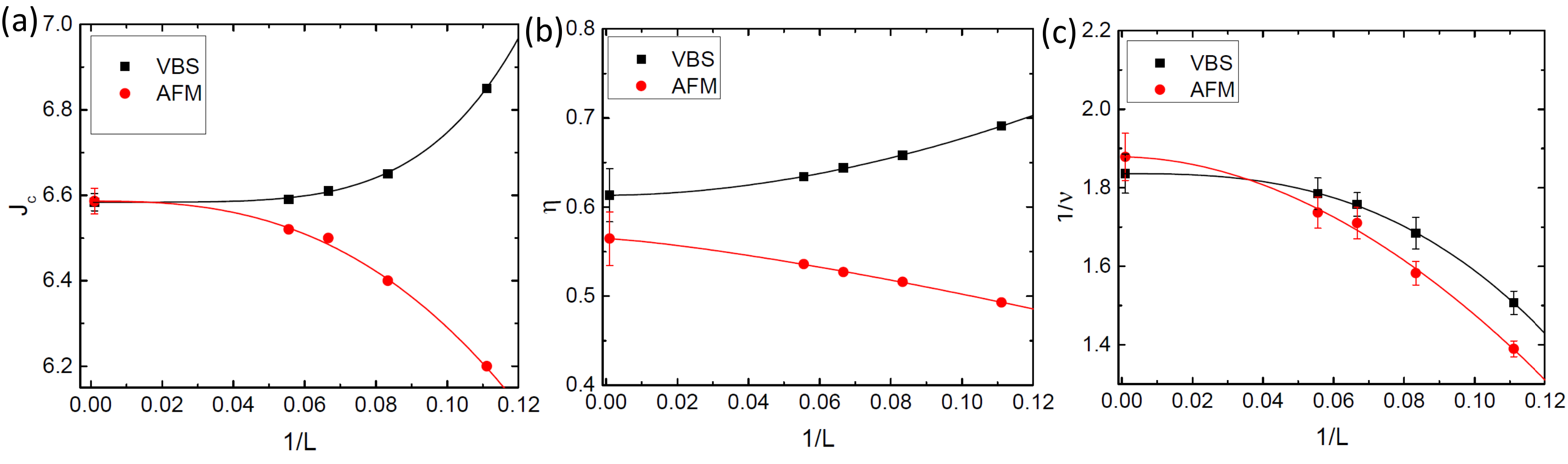}
\caption{ The results of phase transition point and critical exponents for VBS-AFM transition for $U=0.5$. (a) The extrapolation of critical values $J_c$  to the thermodynamic limit for VBS and AFM phase transitions. The largest system size in our simulation is $L=21$. The transition points are $J_{\textrm{VBS}}^c = 6.58\pm0.02$ and $J_{\textrm{AFM}}^c = 6.58\pm0.03$. (b) The extrapolation of anomalous dimension $\eta$ to thermodynamic limit. The fitted result is $\eta_{\textrm{VBS}} = 0.61\pm 0.03$ and $\eta_{\textrm{AFM}} = 0.59\pm 0.02$. (c) The extrapolation of correlation function exponent $\nu$ to the thermodynamic limit. The fitted result is $1/\nu_{\textrm{VBS}} = 1.83\pm0.05$ and $1/\nu_{\textrm{AFM}} = 1.87\pm0.06$ . }
\label{figS1}
\end{figure}

\begin{figure}[t]
\includegraphics[height=3.5cm]{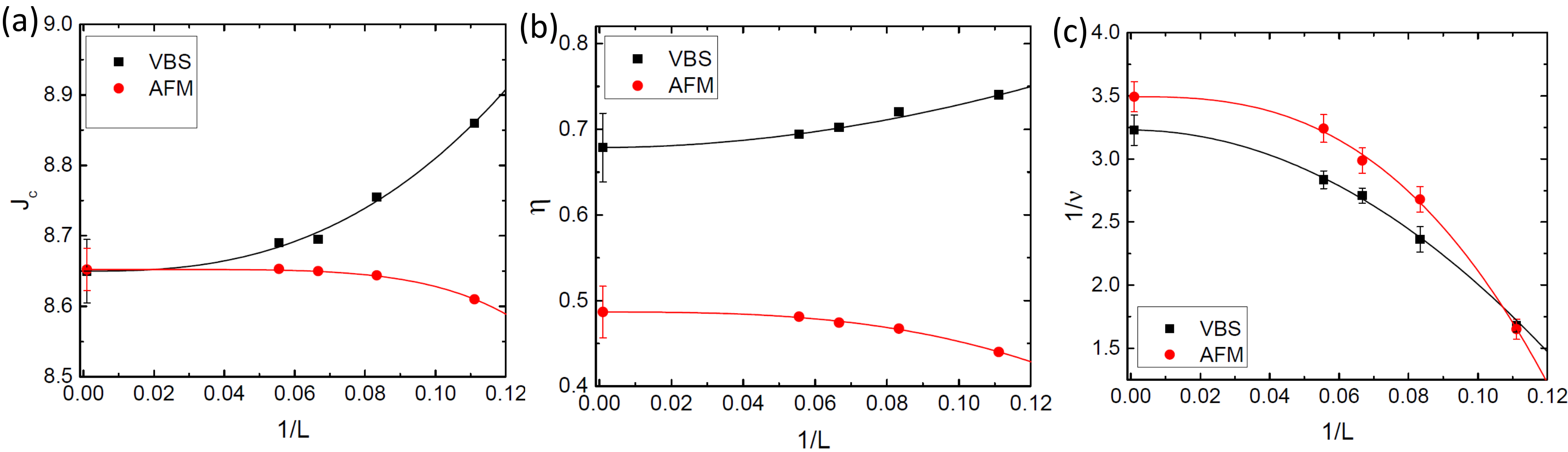}
\caption{ The results of phase transition point and critical exponents for VBS-AFM transition for $U=0$. (a) The extrapolation of critical values $J_c$ to the thermodynamic limit for VBS and AFM phase transition points. The largest system size in our simulation is $L=21$. The transition points are $J_{\textrm{VBS}}^c = 8.65\pm0.05$ and $J_{\textrm{AFM}}^c = 8.65\pm0.03$.  (b) The extrapolation of anomalous dimension $\eta$ to the thermodynamic limit for VBS and AFM order. The fitted result is $\eta_{\textrm{VBS}} = 0.68\pm 0.04$ and $\eta_{\textrm{AFM}} = 0.49 \pm 0.03 $. (c) The extrapolation of correlation function exponent $\nu$ to the thermodynamic limit for VBS and AFM order. The fitted result is $1/\nu_{\textrm{VBS}} = 3.23\pm0.1$ and $1/\nu_{\textrm{AFM}} = 3.49\pm 0.12$. }
\label{figS2}
\end{figure}

\subsection{II. Finite size scaling analysis of critical properties }

\begin{figure}[b]
\includegraphics[height=3.2cm]{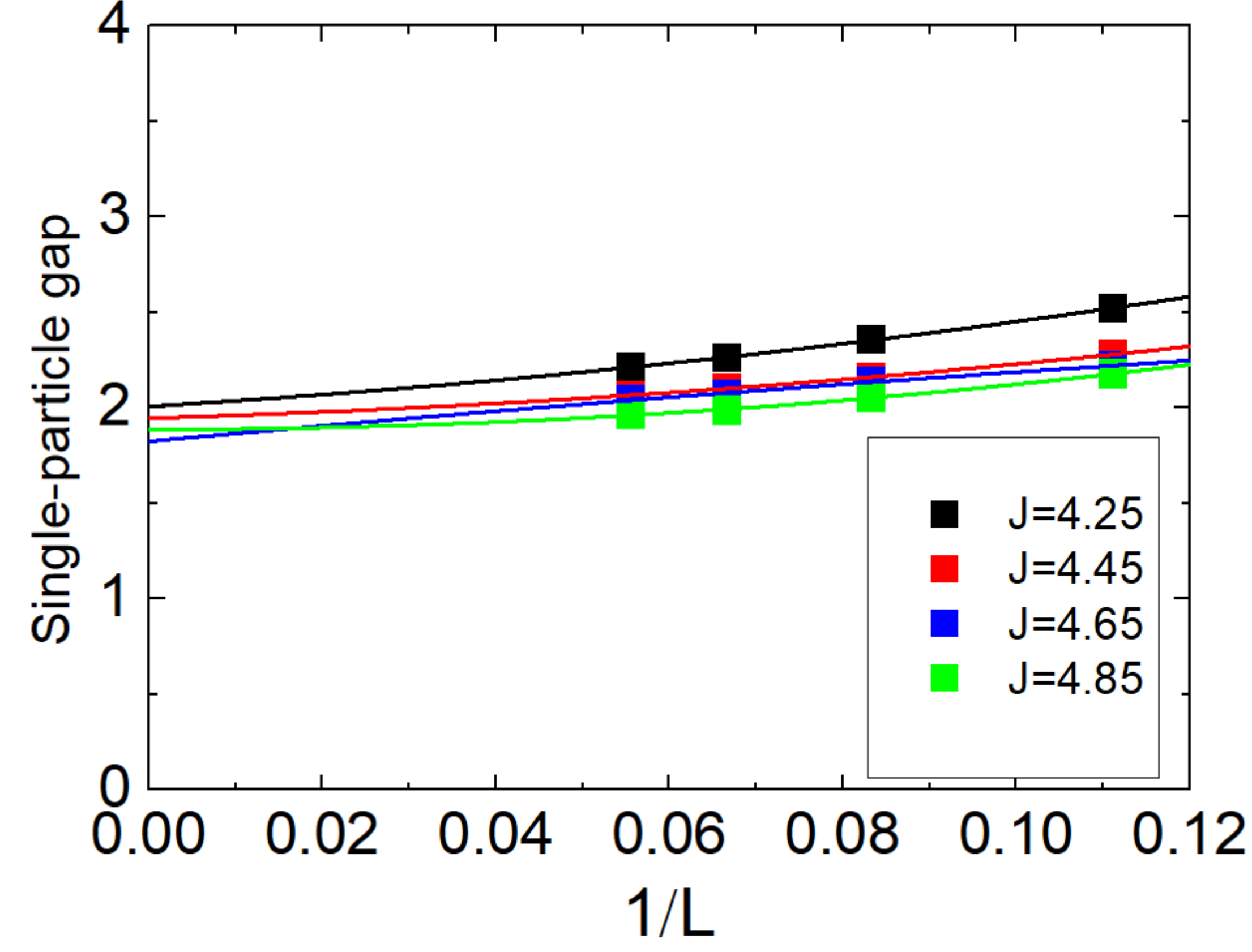}
\caption{ The results of single-particle gap for $U=1.0$ and different values of $J$. The values of single-particle gap for different system sizes are fitted by the second-polynomial function $\Delta(L) = \Delta(L \rightarrow \infty) + a/L + b/L^2$. The results clearly show that across the VBS-AFM transition point, namely $J=4.6$, the single-particle gap is persistent. }
\label{figS3}
\end{figure}

In the present work, we investigate the nature of quantum phase transitions of two ordered phases, the Kekul\'e VBS and AFM, by evaluating their structure factors in the QMC simulations, which are defined as the Fourier transform of correlation function:
\bea
S(\vec{q},L) = \frac{1}{L^4} \sum_{i,j}\avg{\hat{O}_i\hat{O}_j}e^{i(\vec r_i-\vec r_j)\cdot \vec{q}},
\eea
where $\hat O$ represents the VBS or AFM order parameter, $ij$ are site indices, $L$ denote the system size, and $\vec q$ is the crystalline momentum.
For VBS order, the observable is $\hat{O}_i = \sum_{\sigma}(c^\dagger_{i\sigma}c_{i+\hat\delta,\sigma} + H.c)$, where $\hat{\delta}$ labels the direction of nearest-neighbor bond, and the peaked momentum is $\vec{K} = (\pm \frac{4\pi}{3},0)$. For AFM order on honeycomb lattice, the observable is $\hat{O}_i = S_{i,A}^z-S_{i,B}^z$, where $S_{i,A/B}^z = c^\dagger_{i,A/B,\uparrow}c_{i,A/B,\uparrow}-c^\dagger_{i,A/B,\downarrow}c_{i,A/B,\downarrow}$ is spin operator in z-direction on the site of A/B sublattice in unit cell $i$, and the peaked momentum is $\vec{K}=(0,0)$.

The RG invariant ratio, which is the ratio of structure factor defined in maintext, is a powerful tool to determine the phase transition point. In the long-range ordered phase, the RG invariant ratio $R(L)\rightarrow 1$ for $L\rightarrow \infty$, whereas $R(L) \rightarrow 0$ for $L\rightarrow \infty$ in the disordered phase. When system is large enough, the RG-invariant ratio is independent of system size at putative QCP due to the scaling invariance in the long-distance limit. Consequently, the phase transition point can be identified through the crossing point of RG-invariant ratio curve for different system sizes. In order to access the property in the long-distance limit, we perform finite-size scaling of the crossing point of RG-invariant ratio for different system sizes. We identify the transition point $J_c(L)$ in finite system with linear system size $L$ through the crossing point of RG-invariant ratio for $L$ and $L+3$. Then we fit the values of $J_c(L)$ for $L=9,12,15,18$ using the general scaling function $J_c(L) = J_c(L\rightarrow \infty) + \frac{a}{L^b}$. The extrapolated result of $J_c(L\rightarrow \infty)$ indicates the accurate phase transition point in the thermodynamic limit. Employing the finite-size scaling of phase transition point, we obtain the phase boundaries of DSM, VBS, and AFM phases, as shown in \Fig{fig1}.

The critical exponents can also be extracted by structure factor and RG-invariant ratio according to their universal scaling behaviours around QCP. The universal scaling functions describing structure factor at peaked momentum and RG-invariant ratio around QCP are:
\bea
S(\vec{K},L) &=& L^{-(d+z-2+\eta)}F_1((J-J_c)L^{1/\nu},L^{-b_1}) \\
R(L) &=& F_2((J-J_c)L^{1/\nu},L^{-b_2})
\eea
Here, $\vec{K}$ is the peaked momentum of VBS/AFM structure factor. The critical exponent $\eta$ is anomalous dimension and $\nu$ is correlation function exponent. $z$ is dynamical critical exponent which is $z=1$ in our case due to the Dirac physics. The terms $L^{-b_1}$ and $L^{-b_2}$ are subleading finite-size corrections. $F_1$ and $F_2$ are unknown ansatz scaling function. Based on above scaling function, we can extract the values of $\eta$:
\bea
\eta(L) = \frac{1}{\log(\frac{L}{L+3})} \log(\frac{S(\vec{K},L+3)}{S(\vec{K},L)})\mid_{J=J_c} - (d+z-2),
\eea
and the value of $\nu$:
\bea
1/\nu(L) = \frac{1}{\log(\frac{L+3}{L})} \log(\frac{\frac{d}{d J} R(L+3)}{ \frac{d}{d J} R(L)})\mid_{J=J_c}.
\eea
To obtain the critical exponents in the thermodynamic limit, we also perform finite-size scaling by extrapolating the values of $\eta(L)$ and $1/\nu(L)$ obtained in finite systems to $L \rightarrow \infty$. We use the general scaling function to fit the critical exponents: $C(L) = C(L\rightarrow \infty) + \frac{c}{L^b}$, where $C(L) = \eta(L)$ or $1/\nu(L)$. Employing the finite-size scaling analysis, we obtain the critical exponents of the direct VBS-AFM transition for $U=0$, $0.5$, and $1.0$.

\subsection{III. The QMC results of $U=0.5$ and $U=0$ }
Employing the same procedure for the case of $U=1.0$, we obtained the critical values of $J$ and critical exponents for the VBS-AFM phase transition for both $U=0.5$ and $U=0$. The results for $U=0.5$ are shown in \Fig{figS1}. From the finite-size scaling analysis, we can draw the conclusion that the critical values of $J$ for the VBS and AFM transitions are identical. Namely, there is a direct transition between VBS and AFM. Moreover, the extracted critical exponents $\eta$ and $\nu$ for VBS and AFM orders are consistent with each other within error bars. The consistency indicates that the transition between VBS and AFM phases is a single continuous transition. The extracted critical exponents for $U=0.5$ are also consistent with the results for $U=1.0$. For the case of $U=0$, the system also undergoes a direct transition from VBS to AFM phases. Nevertheless, the critical exponents of VBS and AFM transition are different. The extracted critical exponents $\nu$ for VBS and AFM phases are both smaller than $1/3$, which is an important indication of first-order transition in numerical simulation. The results for $U=0$ are shown in \Fig{figS2}.

\subsection{IV. The results of single-particle gap }
We compute the single-particle gap through time-dependent single-particle Green's function, which satisfies the scaling behaviour $c_{\vec{K}}(\tau) c_{\vec{K}}^\dagger(0)\sim e^{-\Delta_{sp} \tau}$ asymptotically when $\tau$ is sufficiently large. Here, $\vec{K}=(\frac{4\pi}{3},0)$ is the momentum of the Dirac point in the Brillouin zone of the honeycomb lattice. The single-particle gaps for different system sizes can thus be extracted from the scaling behaviour of time-dependent Green's function. The value of single-particle gap in the thermodynamic limit is obtained by finite-size scaling analysis using a general second-order polynomial scaling function: $\Delta_{sp}(L) = \Delta_{sp}(L\rightarrow \infty) + \frac{a}{L} + \frac{b}{L^2}$. We plot the results of single-particle gap in the parameter regime $ 4.25<J<4.85$ with fixed value of $U=1.0$. The results clearly show that the single-particle gap remains finite across the transition between VBS and AFM phases. The persistence of finite single-particle gap indicates that the effective field theory describing the VBS-AFM transition is purely bosonic and can be obtained through integrating out gapped fermions.

\begin{figure}[t]
\includegraphics[height=3.5cm]{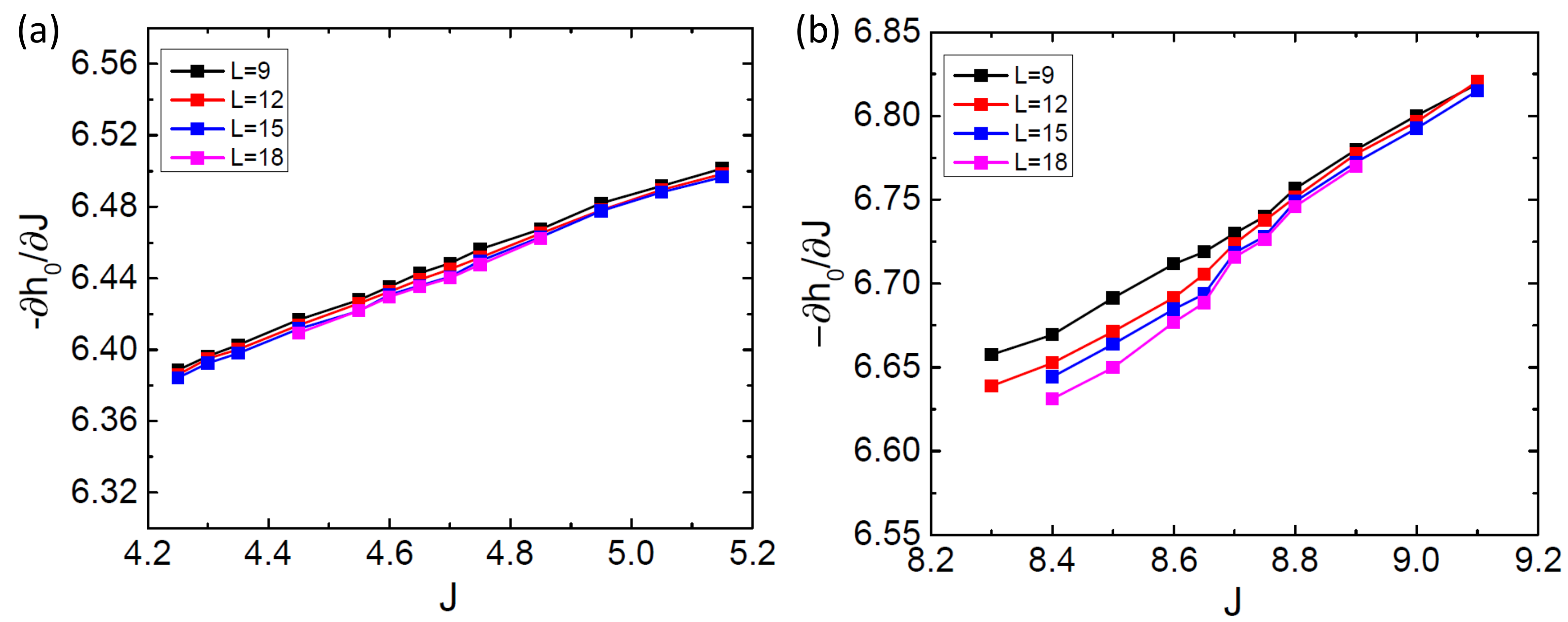}
\caption{(a) The results of first-order derivative of ground-state energy density as a function of $J$ for $U=1.0$ . (b) The results of first-order derivative of ground-state energy density as a function of $J$ for $U=0$. The sharp kink at VBS-AFM phase transition point in the result of $U=0$ is an indication of first-order transition. }
\label{figS4}
\end{figure}

\subsection{V. The results of first-order derivative of ground-state energy}
We compute the first-order derivative of ground-state energy density $E_0$ with respect to $J$ in the regime around VBS-AFM transition:
\bea
\frac{\partial E_0}{\partial J} = -\frac{1}{4L^2} \sum_{\avg{ij},\sigma} \avg{(c^\dagger_{i\sigma}c_{j\sigma} + H.c)^2}.
\eea
The discontinuity of this quantity at the transition in the thermodynamic limit is a hallmark of first-order phase transition. We present the results of first-order derivative of ground-state energy for $U=0$ and $U=1.0$. For $U=1.0$, the tendency of discontinuity is absent with increasing system sizes. In contrast, a sharp kink, which is more pronounced with increasing system sizes, appears at the transition point for the case of $U=0$. The clear distinction between the results of two cases provides convincing evidence that the VBS-AFM transition is continuous for $U=1.0$ while it is first-order for $U=0$.

\begin{figure}[t]
\includegraphics[height=3.5cm]{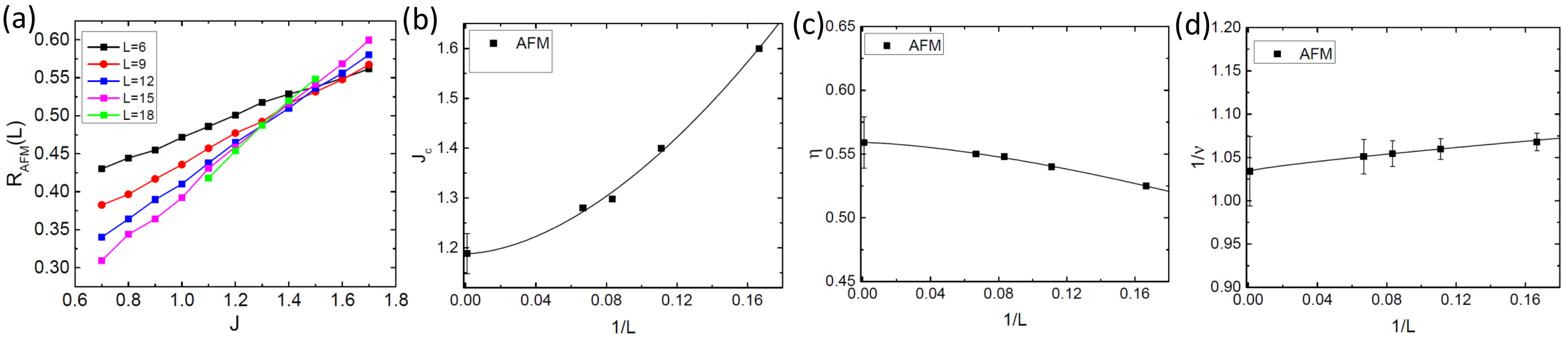}
\caption{(a) The results of RG-invariant ratio for AFM order at $U=3.0$. (b) The extrapolation of critical values $J_c$ for AFM phase transitions to thermodynamic limit. The largest system size in our simulation is $L=18$. The transition point is $J_{\textrm{VBS}}^c = 1.19\pm0.03$. (c) The extrapolation of anomalous dimension $\eta$ to thermodynamic limit. The fitted result is $\eta = 0.56\pm 0.03$. (d) The extrapolation of correlation function exponent $\nu$ to thermodynamic limit. The fitted result is $1/\nu = 1.03\pm0.04 $. }
\label{figS5}
\end{figure}

\subsection{VI. The critical exponents at the DSM to AFM transition}
In addition to the VBS-AFM transition, we also investigate the critical properties of the phase transition between DSM to AFM phases in the model under study. At the DSM-AFM transition, the AFM order-parameter boson condenses and the Dirac fermion opens up a gap due to the non-zero mass terms generated by AFM ordering. The quantum phase transition is described by the Gross-Neveu-Yukawa model and belongs to chiral Heisenberg universality class. When $U>2.0$, the VBS phase does not appear and the ground state is AFM when $J$ is relatively large. Employing the same finite-size scaling procedure, we obtain accurate phase boundary between DSM and AFM phases, as shown in the phase diagram of \Fig{fig1}. Setting $U=3.0$, we perform finite-size scaling and extract the critical exponents $\eta$ and $\nu$ for the DSM-AFM phase transition, the results of which are shown in \Fig{figS5}. The results $\eta = 0.56\pm0.03 $ and $1/\nu = 1.03 \pm0.04  $ are consistent with previous numerical results of chiral Heisenberg universality class.

\subsection{VII. Histogram analysis of Kekul\'e VBS order parameter}
 At the VBS-AFM transition, the continuous $U(1)$ symmetry is predicted to emerge at low energy if the transition is continuous. In order to verify the emergent $U(1)$ symmetry at the VBS-AFM transition point, we employ the histogram technique by evaluating the concurrence probability of real and imaginary parts of the VBS order parameter $P(\Re(\phi_{\textrm{VBS}},\Im(\phi_{\textrm{VBS}}))$. Away from the transition point, the concurrence probability should exhibit the expected $C_3$ symmetry in the histogram. At the VBS-AFM transition point, the concurrence probability should be rotationally invariant if the transition point is a DQCP where the $U(1)$ symmetry emerges. The result of histogram analysis at VBS-AFM transition point is shown in \Fig{fig4}(a). Fixing $U=1.0$ and $J=4.6$, we can clearly observe that the concurrence probability is rotational invariant and continuous $U(1)$ symmetry emerges.
For comparison, we also perform the histogram analysis of Kekul\'e VBS order parameter in the regime away from the VBS-AFM transition. We present the results of VBS histogram for $J=4.2$ and $J=5.1$ in \Fig{figS6}, which clearly show that the concurrence probability of VBS order parameter only exhibits $C_3$ symmetry. The absence of $U(1)$ symmetry away from VBS-AFM transition point verifies that the appearance of $U(1)$ symmetry at the VBS-AFM transition is an emergent phenomenon at DQCP.

\begin{figure}[t]
\includegraphics[height=3.5cm]{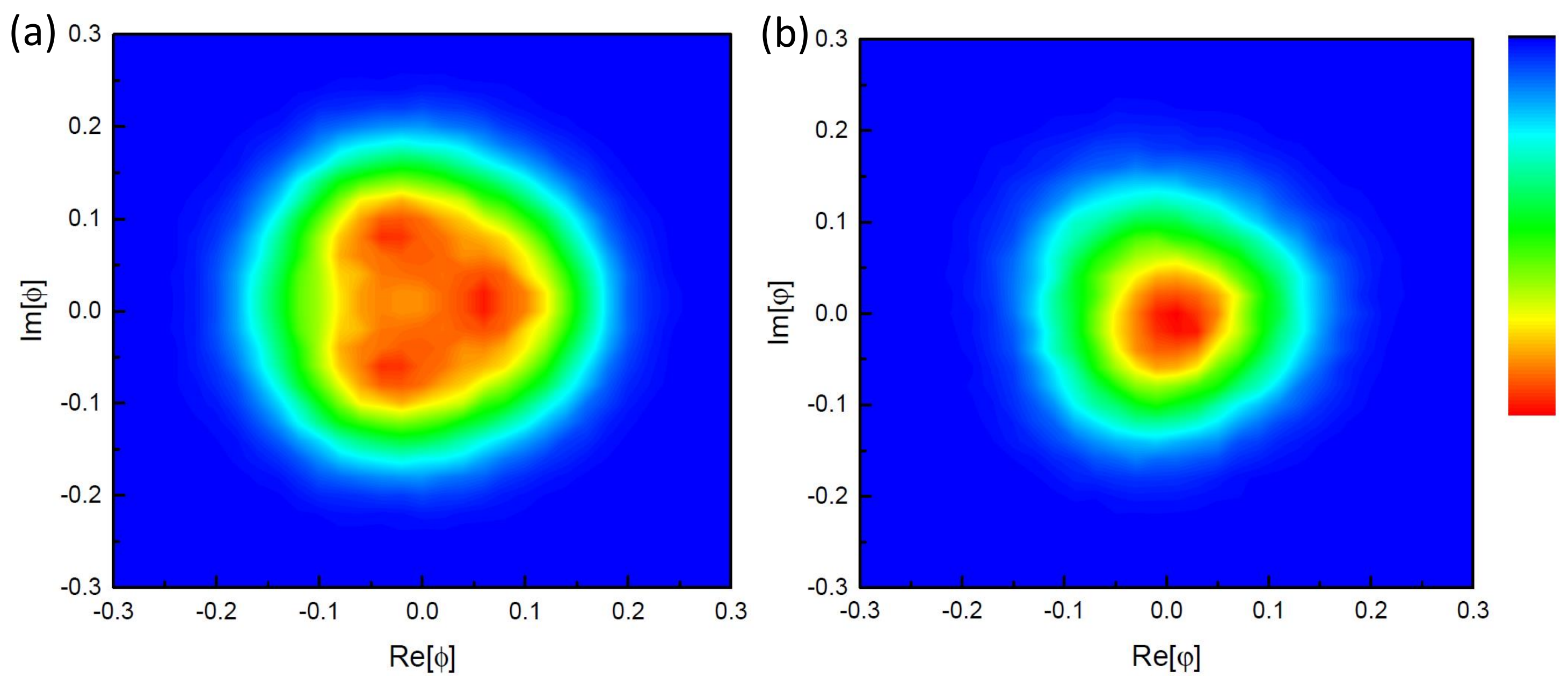}
\caption{The histogram analysis of VBS order away from VBS-AFM transition point. (a) The concurrence probability of VBS order parameter $P(\Re[\phi],\Im[\phi])$ for $L=15$, $U=1.0$ and $J=4.2$. (b) The concurrence probability of VBS order parameter $P(\Re[\phi],\Im[\phi])$ for $L=15$, $U=1.0$ and $J=5.1$. }
\label{figS6}
\end{figure}

\subsection{VIII. Effective non-linear sigma model with one or two WZW terms}
In this section, we shall derive the effective non-linear sigma model describing the VBS-AFM transition at $U=0$, where the model respects an additional spin-particle-hole symmetry, and then discuss its possible connection to the first-order transition. At $U=0$, the spin-particle-hole $Z_2$ symmetry relates the AFM order and pseudospin AFM order. The low-energy noninteracting Dirac Hamiltonian on the honeycomb lattice is given by
\bea
	\mathcal H_0 =\int d^2 p~\psi^\dag (\tau^z \sigma^x s^0 p_x + \tau^0 \sigma^y s^0 p_y) \psi,
\eea
where $\sigma$, $\tau$, $s$ refers to sublattice, valley, and spin degrees of freedom, and $\psi({\vec p})$ denote eight-component electron annihilation operator at momentum ${\vec p}$. Note that we have rescaled the momentum to set Fermi velocity to one. The three-component AFM and two-component VBS orders in terms of low-energy fermion are given by
\bea
 	\vec{\phi}_{\textrm{AFM}} = \psi^\dag( \sigma^z s^x, \sigma^z s^y, \sigma^z s^z) \psi, \quad \quad
	\vec{\phi}_{\textrm{VBS}} = \psi^\dag( \tau^x \sigma^x, \tau^y \sigma^x)\psi,
\eea
respectively. Since we need to consider superconducting (SC) order, we use the Nambu notation, i.e.,
\bea
	\mathcal H_0 = \int d^2p ~\Psi^\dag (\tau^z \sigma^x p_x + \sigma^y \mu^z p_y) \Psi,
\eea
where $\Psi({\bf p}) \equiv (\psi({\bf p}), \psi^\dag(-{\bf p}) )^T$ is the Nambu spinor and $\mu$ are Pauli matrices in Nambu space. Then the pseudospin (SC/CDW) order parameters are given by
\bea
 	\vec{\phi}_{\textrm{pseudospin-AFM}}=(\phi_{\textrm{SC}}^1, \phi_{\textrm{SC}}^2, \phi_{\textrm{CDW}}) = \Psi^\dag(\tau^x s^y \mu^x, \tau^x s^y \mu^y, \sigma^z \mu^z) \Psi.
\eea
Notice the spin-particle-hole $Z_2$ symmetry can relate $\vec \phi_{\textrm{AFM}}$ and $\vec \phi_{\textrm{pseudospin-AFM}}$.

As the symmetry breaking patterns in the two sides of the transition are distinct, it leads to two independent set of mass terms in mean field level. In order to get a genuine deconfined second-order transition, at the transition point there must emerge some kind of enlarged symmetry that can relate these two mass terms in the IR. In the scenario of DQCP at the VBS-AFM transition, SO(5) symmetry that can rotate $\phi_\text{VBS}$ into $\phi_\text{AFM}$ emerges in the IR, which in turn relates two independent mass terms $r_\text{VBS} \sum_{i=1}^2 (\phi_\text{VBS}^i)^2 $ and $r_\text{AFM} \sum_{i=1}^3 (\phi_\text{AFM}^i)^2 $. The reason behind the emergence of SO(5) symmetry is due to the topological WZW term which describes the mapping from the extended spacetime manifold $S^4$ to compact target order-parameter manifold $S^4$. When $U=0$, due to the spin-particle-hole symmetry, not only $\vec \phi_\text{VBS}$ and $\vec \phi_\text{AFM}$ close gap, so does $\vec \phi_\textrm{pseudospin-AFM}$. Now the target manifold is enlarged to an eight-component real boson. A direct compactification of the ground state manifold to $S^7$ with putative SO(8) is unstable because of the relevance of quadratic term as well as the absence of topological terms, i.e. $\pi_4(S^7)=0$. On the other hand, in order to take the advantage of the WZW terms, we can instead compactify the ground state manifold to $M$ given by the following conditions,
\bea\label{condition}
	\sum_{i=1}^2 (\phi_\text{VBS}^i)^2 + \sum_{j=1}^3 (\phi_\text{AFM}^j)^2 &=& 1, \\
	\sum_{i=1}^2 (\phi_\text{VBS}^i)^2 + \sum_{j=1}^3 (\phi_\textrm{pseudospin-AFM}^j)^2 &=& 1.\label{condition2}
\eea
Apparently, there are two submanifolds $S^4 \subset  M$, one can get two WZW terms instead of one, since presumably $\pi_4(M)= Z\times Z$.

To help deriving non-linear sigma model (NLSM), we use the compact notation
\bea
	(\Phi_1, \Phi_2, \Phi_3) \equiv \vec{\phi}_{\textrm{AFM}} , \quad (\bar \Phi_1,\bar \Phi_2 , \bar \Phi_3) \equiv (\phi_{\textrm{SC}}^1, \phi_{\textrm{SC}}^2, \phi_{\textrm{CDW}}), \quad (\Phi_4, \Phi_5) = (\bar \Phi_4, \bar \Phi_5) \equiv \vec{\phi}_{\textrm{VBS}},
\eea
and $\sum_{i=1}^{5} \Phi_i^2 = \sum_{i=1}^{5} \bar \Phi_i^2 =1$. By introducing Gamma notation, $\gamma^0 = \sigma^z$, $\gamma^1= \tau^z \sigma^y$, $\gamma^2= \sigma^x \mu^z$, $\gamma^3= \tau^x \sigma^y \mu^z$, and $\gamma^5= \tau^y \sigma^y$ and $\bar \Psi = \Psi^\dag \gamma^0$, the Lagrangian density of quadratic Dirac fermions reads
\bea
	\mathcal L = \bar \Psi ( \slashed \partial + m \mathcal{V} ) \Psi,
\eea
where $m$ denotes the gap of fermions as fermionic excitations are gaped throughout the VBS and AFM phase, and
\bea	
	\mathcal{V} =  \Phi_1 S^x + \Phi_2 S^y + \Phi_3 S^z + \bar \Phi_1 \Sigma^x + \bar \Phi_2 \Sigma^y + \bar \Phi_3 \Sigma^z  +i \Phi_4 \gamma^3 + i \Phi_5 \gamma^5,
\eea
where ${\bf S} \equiv (s^x \mu^z, s^y,  s^z \mu^z)$ and ${\bf \Sigma} \equiv (\tau^x s^y \mu^x, \tau^x \sigma^y s^y \mu^y, \mu^z)$. Note that $S^i$ and $\Sigma^j$ commutes. Here, the Dirac operator, $\mathcal D \equiv \slashed \partial + m \mathcal{V}$,  has the property
\bea
	\mathcal D^\dag \mathcal D = -\partial^2 + m^2 + m^2 \big(\sum_{i,j=1}^3 \Phi_i \bar\Phi_j \Sigma^i S^j - \Phi_4^2 -\Phi_5^2\big)+ m \slashed \partial \mathcal V,
\eea
from which we can get the inverse of Dirac operator
\bea
	\mathcal D^{-1} = \frac1{-\partial^2 + m^2 + m^2 \big(\sum_{i,j=1}^3 \Phi_i \bar\Phi_j \Sigma^i S^j - \Phi_4^2 -\Phi_5^2\big)+ m \slashed \partial \mathcal V} D^\dag.
\eea
Now we integrate out Dirac fermions to get the effective action of these order parameter, $W = - \Tr \log \mathcal D$. Consider an arbitrary variation of the order parameter,
\bea
	\frac{\delta W}{\delta \mathcal V}= - \Tr \mathcal D^{-1} \delta D = - \Tr \frac1{-\partial^2 + m^2 + m^2 \big(\sum_{i,j=1}^3 \Phi_i \bar\Phi_j \Sigma^i S^j - \Phi_4^2 -\Phi_5^2\big)+ m \slashed \partial \mathcal V} D^\dag \delta D.
\eea
By expanding the operator, i.e.,
\bea
	\frac1{-\partial^2 + m^2 + m^2 \big(\sum_{ij} \Phi_i \bar\Phi_j \Sigma^i S^j - \Phi_4^2 -\Phi_5^2\big)+ m \slashed \partial \mathcal V} = \sum_n \Big[ \frac{m^2 (\sum_{i,j=1}^3 \Phi_i \bar\Phi_j \Sigma^i S^j - \Phi_4^2 -\Phi_5^2)+ m \slashed \partial \mathcal V}{-\partial^2 +m^2} \Big]^n \frac1{-\partial^2 + m^2 }, \nonumber \\
\eea
a direct computation shows two WZW terms emerge at $n=3$ order. Thus, the effective action reads
\bea\label{action}
S =\frac{1}{2g}\int d^3 x\left[(\partial_\mu\Phi^i)^2 + (\partial_\mu\bar \Phi^i)^2\right] + S_\textrm{WZW}[\Phi] + S_\textrm{WZW}[\bar \Phi] + \int d^3 x \{ \lambda[(\Phi^i)^2-1] + \bar \lambda[(\bar \Phi^i)^2-1] \}.
\eea
where $g$ denotes the coupling of the NSLM, and the Legendre fields $\lambda$ and $\bar \lambda$ are introduced to enforce the constraints (\ref{condition}) and (\ref{condition2}) classically. And there are two WZW terms. Each WZW term is given by
\bea
	S_\text{WZW}( A )=  \frac{-2\pi i \epsilon_{abcde}}{\text{Area}(S^4)} \int_0^1 d \xi \int d^3x A_a \partial_\xi A_b \partial_\tau A_c \partial_x A_d \partial_y A_e,
\eea
where $A$ denotes vectors formed by order parameters, and $A(\xi, x)$ is embed into a map such that $A(0, x) = (1, 0,0,0,0)$ and $A(1, x) = A(x)$.

In above derivation we neglect all anisotropy terms and the order parameters have $SO(5) \times SO(5) \times Z_2$ symmetry. However, the UV symmetry of the model is only $SO(3) \times SO(3) \times C_3 \times Z_2$, where two SO(3) symmetries refer to spin and pseudospin rotational symmetries, respectively. $C_3$ is the rotational symmetry of the lattice and $Z_2$ refers to the spin-particle-hole symmetry. Considering only the UV symmetry $SO(3) \times SO(3) \times C_3 \times Z_2$, the following anisotropy terms are allowed:
\bea\label{anisotropy}
    S_\text{anisotropy} = v \sum_{i=4,5} \Phi_i^2 + h  (\Phi_4^3 - 3 \Phi_4 \Phi_5^2),
\eea
where $v$ characterizes the anisotropy between VBS and AFM (pseudospin SC/CDW) orders, and $h$ captures the $C_3$ anisotropy. However, there are issues about Eq.~(\ref{action}) and Eq.~(\ref{anisotropy}). We know the spin-particle-hole symmetry is spontaneously broken by either AFM or SC/CDW ground state. The ground state manifold $M$ that we choose cannot describe the transition classically, because the vector components $\phi_\text{VBS}$ and $\phi_\text{AFM/pseudospin-AFM}$ cannot vanish simultaneously under the conditions Eq.~(\ref{condition}) and (\ref{condition2}). It turns out either there is no topological term at the transition, or quantum fluctuations may bring it out of the classical manifold. If Eq.~(\ref{action}) and Eq.~(\ref{anisotropy}) do describe the transition, it is possible that one term in Eq.~(\ref{anisotropy}) is relevant, such that the field theory has at least two relevant directions, ruining the second order transition.

On the contrary, when $U>0$, there is no symmetry relating AFM order and SC/CDW order, and generically only one of these order will become critical at transition point. For example, when $U>0$ ($U<0$), the AFM (SC/CDW) order becomes critical at the transition point. Since $Z_2$ transformation can map AFM order to SC/CDW order and $U \rightarrow -U$, we consider the case of $U>0$ without  loss of generality. In this case, one can neglect the gaped SC/CDW fluctuation, and the effective action becomes
\bea
    S=\frac1{g} \int d^3x \sum_{i=1}^5 (\partial_\mu \Phi_i)^2  +  S_\text{WZW} (\Phi) + S_\text{anisotropy}.
\eea
According to QMC simulation, both anisotropic terms in Eq.~(\ref{anisotropy}) are rendered irrelevant due to the presence of WZW term $S_\text{WZW} (\Phi)$. And deconfined quantum critical point (DQCP) as well as emergent $SO(5)$ symmetry occur.

Comparing the different scenarios at $U=0$ and $U>0$, we attribute the presence of two WZW terms the possible reason for lack of DQCP at $U=0$. The more detailed understanding of interference effect between topological terms remains an interesting and open question.
\end{widetext}


\begin{thebibliography}{99}
\bibitem{Sachdev-book} S. Sachdev, {\it Quantum Phase Transitions} (Cambridge University Press, Cambridge, Ed. 2, 2011).
\bibitem{Sondhi-RMP} S. L. Sondhi, S. M. Girvin,  J. P. Carini,  and D. Shahar,  
    Rev. Mod. Phys. {\bf 69}, 315 (1997).

\bibitem{Wen-book} X.-G. Wen, {\it Quantum Field Theory of Many-body Systems}, (Oxford
    University Press, New York, 2004).
\bibitem{Fradkin-book} E. Fradkin, {\it  Field Theories of Condensed Matter Physics}, (Cambridge University Press, Cambridge, Ed. 2, 2013).

\bibitem{Senthil-04a} T. Senthil, A. Vishwanath, L. Balents, S. Sachdev, and M. P. A. Fisher, Science {\bf 303}, 1490 (2004).
\bibitem{Senthil-04b} T. Senthil, L. Balents, S. Sachdev, A. Vishwanath, and M. P. A. Fisher, Phys. Rev. B {\bf 70}, 144407 (2004).

\bibitem{Levin-Senthil-04} M. Levin and T. Senthil, Phys. Rev. B {\bf 70}, 220403 (2004).
\bibitem{Motrunich-04} O. I. Motrunich and A. Vishwanath, Phys. Rev. B {\bf 70}, 075104 (2004).
\bibitem{Senthil-06} T. Senthil and M. P. A. Fisher, Phys. Rev. B {\bf 74}, 064405 (2006).
\bibitem{Hu-05} A. Tanaka and X. Hu, Phys. Rev. Lett. {\bf 95}, 036402 (2005).
\bibitem{Lee-10} P. Ghaemi, S. Ryu and D.-H. Lee, Phys. Rev. B {\bf 81}, 081403 (2010).
\bibitem{Jian-18} C.-M. Jian, A. Thomson, A. Rasmussen, Z. Bi and C. Xu, Phys. Rev. B {\bf 97}, 195115 (2018).
\bibitem{You-18} Y.-Z. You, Y.-C. He, C. Xu and A. Vishwanath, Phys. Rev. X {\bf 8}, 011026 (2018).
\bibitem{Max-18} M. A. Metlitski and R. Thorngren, Phys. Rev. B {\bf 98}, 085140 (2018).
\bibitem{Motrunich-19a} S. Jiang and O. Motrunich, Phys. Rev. B {\bf 99}, 075103 (2019).

\bibitem{Sandvik-07} A. W. Sandvik, Phys. Rev. Lett. {\bf 98}, 227202 (2007).
\bibitem{Kaul-08} R. G. Melko and R. K. Kaul, Phys. Rev. Lett. {\bf 100}, 017203 (2008).
\bibitem{Alet-13} S. Pujari, K. Damle, and F. Alet, Phys. Rev. Lett. {\bf 111}, 087203 (2013).
\bibitem{Guo-16} H. Shao, W. Guo, and A. W. Sandvik, Science {\bf 352}, 213 (2016).
\bibitem{Kaul-132} R. K. Kaul, R. G. Melko, and A. W. Sandvik, Annu. Rev. Condens. Matter Phys. {\bf 4}, 179 (2013).
\bibitem{Sandvik-09} J. Lou, A. W. Sandvik, and N. Kawashima, Phys. Rev. B 80, 180414(R)(2009).
\bibitem{Wiese-08} F.-J. Jiang, M. Nyfeler, S. Chandrasekharan, and U.-J. Wiese, J. Stat. Mech. P02009 (2008).
\bibitem{Prokofev-08} A. B. Kuklov, M. Matsumoto, N. V. Prokofev, B. V. Svistunov, and M. Troyer, Phys. Rev. Lett. {\bf 101}, 050405 (2008).
\bibitem{Kaul-12} R. K. Kaul and A. W. Sandvik, Phys. Rev. Lett. {\bf 108}, 137201 (2012).
\bibitem{Xiang-12} H. H. Zhao, C. Xu, Q. N. Chen, Z. C. Wei, M. P. Qin, G. M. Zhang, and T. Xiang, Phys. Rev. B {\bf 85}, 134416 (2012).
\bibitem{Kaul-13} M. S. Block, R. G. Melko, and R. K. Kaul, Phys. Rev. Lett. {\bf 111}, 137202 (2013).
\bibitem{FWang-15} F. Wang, S. A. Kivelson, and D.-H. Lee, Nature Physics {\bf 11}, 959 (2015).
\bibitem{Meng-18a} N. Ma, G.-Y. Sun, Y.-Z. You, C. Xu, A. Vishwanath, A. W. Sandvik, and Z. Y. Meng, Phys. Rev. B {\bf 98}, 174421 (2018).
\bibitem{Meng-18b} N. Ma, Y.-Z. You and Z. Y. Meng,  arXiv:1811.08823 (2018).
\bibitem{Pollmann-18} X.-F. Zhang, Y.-C. He, S. Eggert, R. Moessner and F. Pollmann, Phys. Rev. Lett. {\bf 120}, 115702 (2018).
\bibitem{Gazit-18} S. Gazit, F. F. Assaad, S. Sachdev, A. Vishwanath, and C. Wang, Proc. Natl. Acad. Sci. {\bf 115}, E6987 (2018).
\bibitem{Motrunich-19b} B. Roberts, Shenghan Jiang and O. Motrunich, arXiv:1904.00010 (2019).
\bibitem{Xiang-19} R.-Z. Huang, D.-C. Lu, Y.-Z. You, Z. Y. Meng and T. Xiang, arXiv:1904.00021 (2019).
\bibitem{Sreejith-15} G. J. Sreejith and S. Powell, Phys. Rev. B {\bf 92}, 184413 (2015).
\bibitem{Harada-13} K. Harada, T. Suzuki, T. Okubo, H. Matsuo, J. Lou, H. Watanabe, S. Todo, and N. Kawashima, Phys. Rev. B {\bf 88}, 220408(R)(2013).
\bibitem{Alet-15} S. Pujari, F. Alet, and K. Damle, Phys. Rev. B {\bf 91}, 104411(2015).
\bibitem{Prokofev-13} K. Chen, Y. Huang, Y. Deng, A. B. Kuklov, N. V. Prokofev, and B. V. Svistunov, Phys. Rev. Lett. {\bf 110}, 185701 (2013).
\bibitem{Bartosch-13} L. Bartosch, Phys. Rev. B {\bf 88}, 195140 (2013).
\bibitem{Wen-16} L. Wang, Z.-C. Gu, F. Verstraete, and X.-G. Wen, Phys. Rev. B {\bf 94}, 075143 (2016).

\bibitem{Haldane-88} F. D. M. Haldane, Phys. Rev. Lett. {\bf 61}, 1029 (1988).
\bibitem{Read-Sachdev-89} N. Read and S. Sachdev, Phys. Rev. Lett. {\bf 62}, 1694 (1989).

\bibitem{Wang-17}  C. Wang, A. Nahum, M. A. Metlitski, C. Xu, and T. Senthil, Phys. Rev. X {\bf 7}, 031051 (2017).

\bibitem{Senthil-15} C. Wang and T. Senthil, Phys. Rev. X {\bf 5}, 041031 (2015).
\bibitem{Metlitski-16} M. A. Metlitski and A. Vishwanath, Phys. Rev. B {\bf 93}, 245151 (2016).
\bibitem{Wang-16} N. Seiberg, T. Senthil, C. Wang, and E. Witten, Ann. Phys. {\bf 374}, 395 (2016)
\bibitem{Meng-17} Y. Q. Qin, Y.-Y. He, Y.-Z. You, Z.-Y. Lu, A. Sen, A. W. Sandvik, C. Xu, and Z. Y. Meng, Phys. Rev. X {\bf 7}, 031052 (2017).
\bibitem{Cenke-18} T. Senthil, D. T. Son, C. Wang and C. Xu, arXiv:1810.05174 (2018).

\bibitem{Zhang-97} S.-C. Zhang, Science {\bf 275}, 1089 (1997). 

\bibitem{Bootstrap-16} Y. Nakayama and T. Ohtsuki, Phys. Rev. Lett. {\bf 117}, 131601 (2016).
\bibitem{Bootstrap-RMP19} D. Poland, S. Rychkov, and A. Vichi, Rev. Mod. Phys. {\bf 91}, 15002 (2019).

\bibitem{Nahum-15a} A. Nahum, P. Serna, J. T. Chalker, M. Ortuno and A. M. Somoza, Phys. Rev. Lett. {\bf 115}, 267203 (2015).
\bibitem{Nahum-15b} A. Nahum, J. T. Chalker, P. Serna, M. Ortuno, and A. M. Somoza, Phys. Rev. X {\bf 5}, 041048 (2015).

\bibitem{Troyer-05} M. Troyer and U. J. Wiese, Phys. Rev. Lett. {\bf 94}, 170201 (2005).
\bibitem{Wu-05} C. Wu and S.-C. Zhang, Phys. Rev. B {\bf 71} 155115 (2005).
\bibitem{LJY-15}  Z.-X. Li, Y.-F. Jiang, and H. Yao, Phys. Rev. B {\bf 91}, 241117 (2015).
\bibitem{LJY-16}  Z.-X. Li, Y.-F. Jiang, and H. Yao, Phys. Rev. Lett. {\bf 117}, 267002 (2016).
\bibitem{Wang-15} L. Wang, Y.-H. Liu, M. Iazzi, M. Troyer, and G. Harcos, Phys. Rev. Lett. {\bf 115}, 250601 (2015).
\bibitem{Xiang-16} Z.-C. Wei, C. Wu, Y. Li, S. Zhang, and T. Xiang, Phys. Rev. Lett. {\bf 116}, 250601 (2016).
\bibitem{LY-18} Z.-X. Li and H. Yao, Annual Review of Condensed Matter Physics {\bf 10} 337 (2019).

\bibitem{BSS-81} R. Blankenbecler, D. J. Scalapino, and R. L. Sugar, Phys.Rev. D {\bf 24}, 2278 (1981).
\bibitem{Cerperley-86} D. Cerperley and B. Alder, Science {\bf 231}, 555 (1986).
\bibitem{Sorella-89} S. Sorella, S. Baroni, R. Car, and M. Parrinello, Europhys. Lett.  {\bf 8}, 663 (1989).
\bibitem{White-89} S. R. White, D. J. Scalapino, R. L. Sugar, E. Y. Loh, J. E. Gubernatis, and R. T. Scalettar, Phys.  Rev. B  {\bf 40} 506 (1989).
\bibitem{Assaad-05}  F. F. Assaad and H. G. Evertz, in Computational Many-Particle Physics, Lecture Notes in Physics Vol. 739 (Springer, Berlin, Heidelberg, 2008), pp. 277-356.
\bibitem{Sandvik-98} A. W. Sandvik, Phys. Rev. B {\bf 57}, 10287 (1998).
\bibitem{Prokofev-98} N. Prokofev, B. Svistunov, and I. Tupitsyn, Phys. Lett. A {\bf 238}, 253 (1998).
\bibitem{Gull-11} E. Gull, A. J. Millis, A. I. Lichtenstein, A. N. Rubtsov, M. Troyer, and P. Werner, Rev. Mod. Phys. {\bf 83}, 349 (2011).

\bibitem{Gutierrez-16} C. Gutierrez {\it et al.}, Nature Physics {\bf 12}, 950 (2016).

\bibitem{SSH} W. P. Su, J. R. Schrieffer, and A. J. Heeger, Phys. Rev. Lett. {\bf 42}, 1698 (1979).
\bibitem{Fradkin-83} E. Fradkin and J. E. Hirsch, Phys. Rev. B {\bf 27}, 1680 (1983).
\bibitem{Assaad-SSH} M. Weber, F. F. Assaad, and M. Hohenadler, Phys. Rev. B {\bf 91}, 245147 (2015).

\bibitem{LJJY-17} Z.-X. Li, Y.-F. Jiang, S.-K. Jian and H. Yao, Nature Communications {\bf 8}, 314 (2017).


\bibitem{Jian-17a} S.-K. Jian and H. Yao, Phys. Rev. B {\bf 96}, 195162 (2017).

\bibitem{Herbut-17} L. Classen, I. F. Herbut and M. M. Scherer, Phys. Rev. B {\bf 96}, 115132 (2017).

\bibitem{Herbut-06} I. F. Herbut, Phys. Rev. Lett. {\bf 97}, 146401 (2006).
\bibitem{Assaad-13} F. F. Assaad and I. F. Herbut, Phys. Rev. X {\bf 3}, 031010 (2013).
\bibitem{Sorella-16} Y. Otsuka, S. Yunoki, and S. Sorella, Phys. Rev. X {\bf 6}, 011029 (2016).
\bibitem{Assaad-15} F. P. Toldin, M. Hohenadler, F. F. Assaad, and I. F. Herbut, Phys. Rev. B {\bf 91}, 165108 (2015).
\bibitem{Herbut-17b} N. Zerf, L. N. Mihaila, P. Marquard, I. F. Herbut, and M. M. Scherer, Phys. Rev. D {\bf 96}, 096010 (2017).
    
\bibitem{Guo-18} Y. Liu, Z. Wang, T. Sato, M. Hohenadler, C. Wang, W. Guo and F. F. Assaad, arXiv:1811.02583 (2018).
\bibitem{Zaletel-18} M. Ippoliti, R. S. K. Mong, F. F. Assaad and M. P. Zaletel, Phys. Rev. B {\bf 98}, 235108 (2018).

\bibitem{Herbut-18} L. Janssen, I. F. Herbut, and M. M. Scherer, Phys. Rev. B {\bf97}, 041117 (2018). 
\bibitem{Roy-18} B. Roy, P. Goswami and V. Juricic, Phys. Rev. B {\bf 97}, 205117 (2018).
\bibitem{Assaad-17b} T. Sato, M. Hohenadler and F. F. Assaad, Phys. Rev. Lett. {\bf 119}, 197203 (2017).

\bibitem{Yang-90} C. N. Yang and S.-C. Zhang, Mod. Phys. Lett. B {\bf 4}, 759 (1990).
\bibitem{Zhang-90} S.-C. Zhang, Phys. Rev. Lett. {\bf 65}, 120 (1990).

\bibitem{Kivelson-15} E. Fradkin, S. A. Kivelson, and J. M. Tranquada, Rev. Mod. Phys. {\bf 87}, 457 (2015).

\bibitem{Kaul-18} J. Wildeboer, J. D'Emidio and R. K. Kaul, arXiv:1808.04731 (2018).
\bibitem{Sandvik-18} B. Zhao, P. Weinberg, A. W. Sandvik, arXiv:1804.07115 (2018).
\end{thebibliography}
\end{document}